\documentclass{optica-article}
\usepackage{subcaption}
\usepackage{graphicx}
\usepackage{hyperref} % Assurez-vous d'inclure ce package
\usepackage{orcidlink} % Permet d'afficher le logo ORCID

\journal{opticajournal} % for journals or Optica Open

\articletype{Research Article}

\usepackage{lineno}
\begin{document}

\title{An all-fibred, telecom technology compatible, room temperature, single-photon source}

\author{Nathan Lecaron~\authormark{1}\orcidlink{0009-0001-3805-1618}, Max Meunier~\authormark{2}, Grégory Sauder~\authormark{1}\orcidlink{0000-0001-8423-547X}, 
Romain Dalidet~\authormark{1}\orcidlink{0009-0007-2755-4547}, Yoann Pelet~\authormark{1}, Sébastien Tanzilli~\authormark{1}\orcidlink{0000-0003-4030-5821}, 
Jes\'us Z\'u\~niga-Pérez~\authormark{2,3}\orcidlink{0000-0002-7154-641X} Olivier Alibart~\authormark{1,*}\orcidlink{0000-0003-4404-4067}}

\address{\authormark{1}Université Côte d'Azur, CNRS, Institut de Physique de Nice, 17 rue Julien Lauprêtre, 06200 NICE, France\\
\authormark{2}Majulab International Joint Research Unit UMI 3654, CNRS, Université Côte d’Azûr, Sorbonne Université, National
University of Singapore, Nanyang Technological University, Singapore, Singapore\\
\authormark{3}Université Côte d’Azur, CNRS, Centre de Recherche sur l’Hétéro Epitaxie
et ses Applications (CRHEA), Rue Bernard Gregory, 06560 Valbonne, France}

\email{\authormark{*}olivier.alibart@univ-cotedazur.fr} %% email address is required; see note below about the corresponding author designation

% use {asbstract*} to suppress the copyright line. Copyright information will be added in production

\begin{abstract*} 
Single photon sources are essential building blocks for fundamental quantum optics but also for quantum information networks. Their widespread is currently hindered by unpractical features, such as operation at cryogenic temperature and emission wavelength lying outside telecom windows. Taking advantage of telecom technology and point defects in GaN crystals, we present, for the first time, the development of a fully-fibred source of single photons operating at room temperature, emitting photons in the telecom O-band and fulfilling the standards of telecom photonics. We characterise an emitter producting single photons at the wavelength of 1292\,nm, a spectral broadening compatible with CWDM channels of 13\,nm, and a brightness of 25\,kcps per mW of pump power. The source shows a signal-to-noise ratio of 16.5 and a purity $g^{(2)}(0) = (5.9 \pm 0.5) \times 10^{-2}$ at room temperature, showing high potential for being integrated transportable quantum cryptography devices.

\end{abstract*}

%%%%%%%%%%%%%%%%%%%%%%%%%%  body  %%%%%%%%%%%%%%%%%%%%%%%%%%
\section{Introduction}

Quantum information technologies have drawn significant attention due to their potential to revolutionize information processing and quantum communication. They include higher resolution and sensitive measurements of physical parameters~\cite{giovannetti2004quantum}, but also more efficient simulations of physical systems~\cite{georgescu2014quantum} and innovative computational tasks~\cite{nielsen2010quantum}. Quantum communication also stands as a central pillar since it enables connecting those technologies for enhanced performances~\cite{kimble2008quantum}, but also offers a more direct application through quantum key distribution (QKD) for sharing secret keys to encrypt sensitive data~\cite{gisin2002quantum, scarani2009security}.

Whatever the intended application, entangled photon pairs and single photon sources emitting at telecom wavelengths are crucial for various quantum communication tasks. From elementary QKD protocols such as BB84 or BBM92~\cite{bennett1984quantum,bennett1992quantum} to more advanced tasks, such as quantum teleportation and entanglement swapping, underpinning quantum internet networks~\cite{teleport1,teleport2}, compact, reliable, and fiber-coupled non-classical photonic sources are needed.

Spontaneous parametric down-conversion in integrated waveguides is a leading candidate for generating such entangled photon pairs due to its industrial maturity, high photon quality, and compatibility with room-temperature operation~\cite{review_IO}. Unfortunately, there is no equivalent situation concerning single-photon sources (SPS).

Advances in semiconductor growth techniques during the last thirty years, in particular molecular beam epitaxy and metalorganic chemical vapor deposition, have indeed enabled the fabrication of tailored semiconductor quantum dots coupled to microcavities, showing unprecedented efficiencies and overall performances. By selecting suitable materials for both the emitting region and the surrounding barrier, it is possible to engineer a two-level system that emits photons within a specific frequency range. The archetypical examples are InGaAs/GaAs quantum dots emitting around 925 nm, which are now even commercially available~\cite{hsieh2005growth,somaschi2016near}, or InAs/InP quantum dots emitting at telecom wavelength~\cite{SPS_telecom}. These devices have facilitated rapid progress in the field of photonic quantum computing~\cite{SPS_computing,SPS_computing2} and quantum communication~\cite{review_SPS_com}. Despite tremendous efforts to offer fiber-coupled devices~\cite{margaria2024}, remaining obstacles to the widespread adoption of QD-based single-photon sources are the need to operate them at cryogenic temperatures around a few Kelvin~\cite{miyazawa2016single}.

Point defects in crystals, so-called colour centres, are another example of SPS in a solid-state environment. These material defects, which can be missing atoms or foreign element inclusions in the otherwise perfectly regular crystal lattice, locally alter the semiconductor structure and introduce localized energy levels within the crystal forbidden bandgap. The most representative case is the nitrogen-vacancy (NV) center in diamond~\cite{kurtsiefer2000stable}, but a plethora of point defects acting as SPS have been identified in a large variety of semiconductors. The material matrices include narrow-bandgap semiconductors, such as silicon~\cite{udvarhelyi2021identification}, wide-bandgap semiconductors, such as SiC~\cite{bathen2021manipulating}, ZnO~\cite{jungwirth2016polarization}, GaN~\cite{zhou2018room}, or AlN~\cite{shi2021point}, as well as insulators such as BN~\cite{grosso2017tunable}. There exist a few room-temperature SPSs whose emission wavelengths are not suited for existing fiber networks, and their potential for integration in fiber-coupled optical circuits is not trivial~\cite{Beveratos2002,Bounouar2012,Holmes2014,Dideriksen2021}.

GaN is, behind silicon, the second most used semiconductor in terms of business share and one can take full advantage of the technological know-how developed so far for solid-state lighting and high-power/high-frequency electronics. Electrical injection as well as photonic building blocks (e.g., Air/GaN distributed Bragg reflectors, optical waveguides, input- and output-couplers, photonic crystals, etc.) already exist in GaN worldwide. More recently, the ability of GaN's defect points to operate as SPS at room temperature~\cite{zhou2018room,arakawa2020progress} and to be seamlessly integrated into photonic structures~\cite{meunier2023telecom}, have positioned this material as a promising contender in the field.

In this article, we present a state-of-the-art performance GaN-based SPS that operates at room temperature using only standard telecommunication technology components. The ease of implementation, the stability of the fully-fibred setup, which will be quantitatively assessed, and the compatibility with off-the-shelf telecom components make this SPS a promising quantum system for real-field QKD protocols.
%--------------------------------------------------------------------
\section{Experimental setup}

\begin{figure*}
    \centering
    \includegraphics[width=\textwidth]{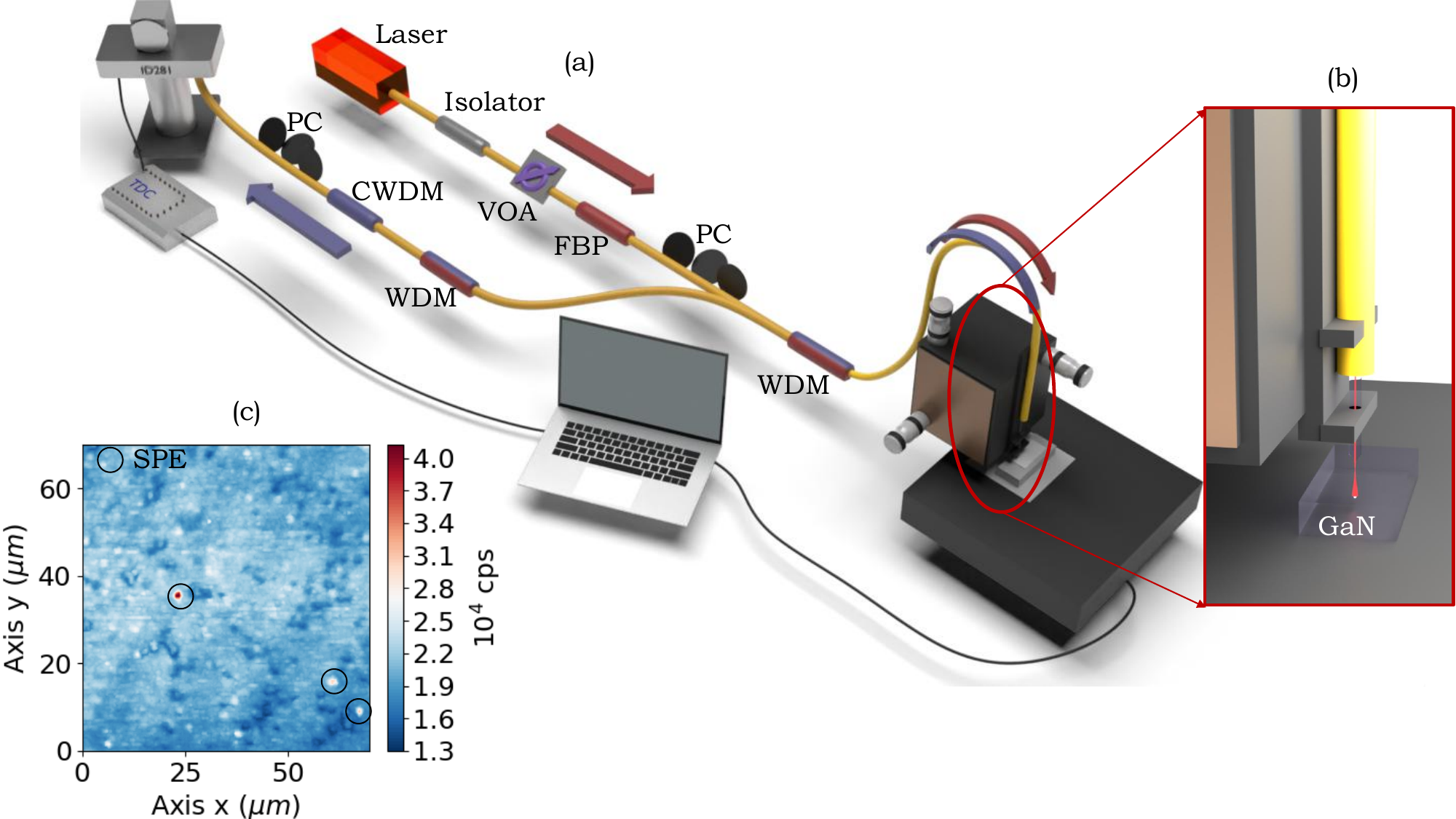}
    \caption{\centering Sketch of fully fibred single photon device operating at telecom wavelength. (a) Scheme of the experimental setup for collecting single photon from GaN samples. We use a CW laser at 976\,nm and a standard 980-1310\,nm WDM to separate the pump photon (red arrow) from the singles photons (blue arrow). (b) Zoomed view the microlensed fibre pointing at the surface of the GaN sample. (c) Intensity map of the GaN sample in the X and Y directions, showing the locations of single photon emitters (SPEs). BPF : band pass filter, ID281 SNSPD : superconducting nanowire single-photon and TDC : Time to digital converter WDM :  wavelength division multiplex, CWDM : Coarse wavelength division multiplex and PC : polarization controller. }
    \label{setup}
\end{figure*}

As shown in figure~\ref{setup}a, the SPS is excited using a continuous-wave (CW) laser at 976\,nm delivering an average output power of about 2\,mW to the sample. The pump beam passes through a 1\,nm amplified spontaneaous emission (ASE) cleaning band-pass filter (BPF) and a wavelength division multiplexer (WDM) in the 980-1310\,nm configuration. The pumping laser is focalised onto the GaN sample by use of a microlensed fibre having a numerical aperture of 0.42. For positioning adjustment, the fibre is mounted on an ultra-stable nanopositioning stage while the sample is installed on a piezo-stage (Jena piezo), which enables to scan the sample surface and locate adequate defect points. This setup is designed to optimize mechanical stability: the control stage allows keeping the mechanical position drift below 300\,nm/h, allowing us to carry out measurements over several hours without requiring active position correction.

After optical excitation of an emitter, single photons are collected by the same microlensed fibre (figure~\ref{setup}b), and routed to the 1310\,nm port of the WDM. Three additional WDMs are placed immediately downstream to achieve the required attenuation (120\,dB for the pump photons). The single-photons are finally detected by a superconducting nanowire single photon detector (SNSPD) showing 90\% efficiency and a dark count rate of 70\,Hz.\\

\section{Searching for efficient emitters and setup stability}

\subsection{Emitter localization}

To identify the position of single-photons emitters (SPEs) we first perform 100\,$\mu$m  $\times$ 100\,$\mu$m photoluminescence spatially-resolved scans. Typically, for the current GaN-on-sapphire thin film, about ten emitters per scan are identified. Figure~\ref{setup}c shows part of the scanned surface (60\,$\mu$m $\times$70\,$\mu$m) where we can see 3 SPSs emitting in the O-band.

To select an optimal emitter compatible with all-fibred system and off-shell telecom components, the following criteria complying with ITU standards were considered: the emission wavelength needs to be centred at one of the standard coarse wavelength division multiplexing (CWDM) channels (i.e. 1271\,nm, 1291\,nm, 1311\,nm, 1331\,nm, or 1351\,nm), and the SPS spectral full-width at half maximum (FWHM) should be narrower than the 13\,nm standard of CDWM channels bandwidth. 
Among the various identified emitters, the measured spectral bandwidths indicate FWHM statistics ranging from a few nanometres to 50\,nm~\cite{zhou2018room,meunier2023etude}. We therefore apply a strict selection among the emitters, showing maximum overlap between the CDWM channels and the measured spectra.

\begin{figure}
    \centering
        \includegraphics[width=\linewidth]{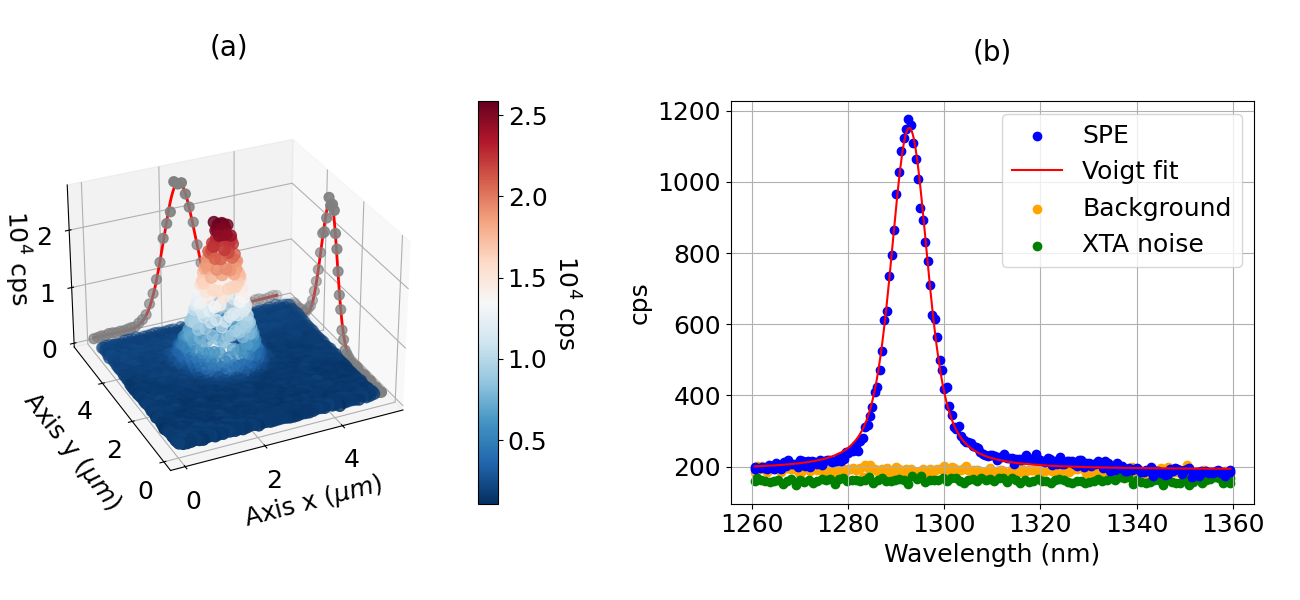}
    \caption{\centering Optical characterisation of an emitter. (a) Spatially-resolved photoluminescence (PL) scan, with the intensity represented along the Z axis and depicted through a color scale. The scan was performed with a pump power of 1 mW and a step size of 100\,nm. (b) PL spectrum of a SPS (blue dots) with a 10 s integration time per point, a pump power of 0.5\,mW, a step size of 500\,pm, and a sampling interval of 900\,pm. The red line is the Voigt fit of the SPS PL emission. The green dots represent the device noise (XTA) when the laser is off and the orange dots represent a spectrum measured in a GaN location where there is no SPS, with the pump laser on.}
    \label{dot_spectre}
\end{figure}

We selected a particular emitter whose spatial intensity distribution and spectral properties are shown in figure~\ref{dot_spectre}. In figure~\ref{dot_spectre}b the blue curve represents the emitter's spectrum, the green one represents thespectrometer (XTA) detection noise, and the orange one corresponds to the optical noise by measuring the surrounding region of GaN without the emitter. The measurement shows a negligible GaN background emission within the O-band. Finally, the red line represents a Voigt fit used to determine the spectral FWHM of the emitter, which displays a central wavelength of 1292\,nm and a bandwidth of $(9.8\pm 0.3)$\,nm. These characteristics meet all the above-mentioned conditions.

Note that while this GaN thin film displays a very low emission in the O-band (figure~\ref{dot_spectre}b), it has a non-negligible optical background when integrated over the WDM spectral windows of 100 nm. This is why we use a 13\,nm spectral filter (CWDM) to increase the emitter signal-to-noise ratio from 2.6 (figure~\ref{setup}b) to 16.5 (figure~\ref{dot_spectre}b). The use of a 13\,nm filter reduces the noise level from 20\,kHz to 1.5\,kHz, while the detection of single photons is only reduced by a factor 1.2, due to some residual filter loss. Under these experimental conditions, we obtained a  brightness of 25\,kcps per mW of pump power while emission intensity saturation is observed at about 5\,mW (figure~\ref{power}b), indicating that the SPS could be operated at a maximal external brightness of around 60\,kHz.

\subsection{Mechanical stability}
\begin{figure}
    \centering
    \includegraphics[width=\linewidth]{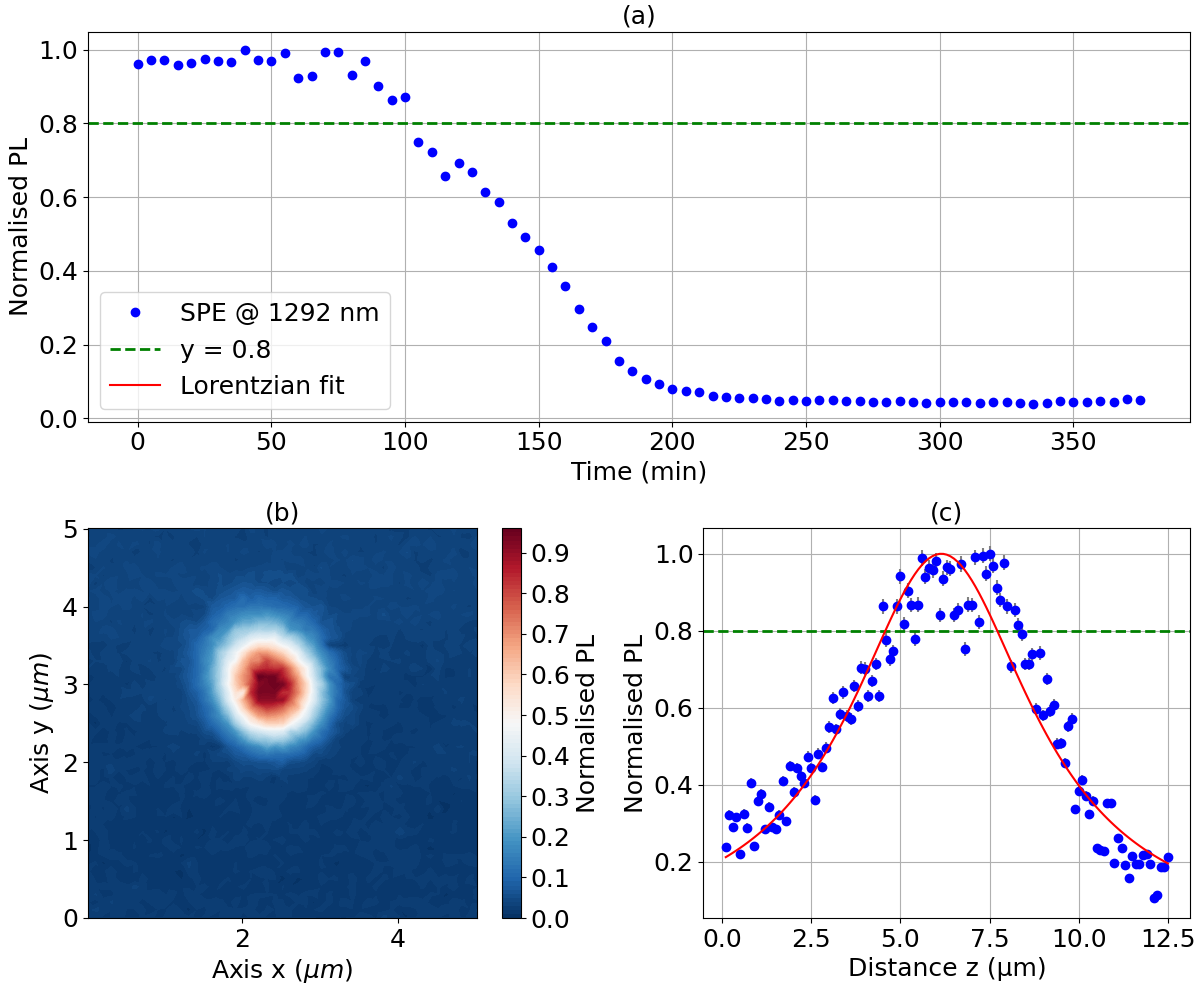}
    \caption{\centering Classical characterisation of the performance of the experimental set-up. The data have been graphically normalized to intensity (cps). The data are represented by blue dots, the fit by a green line and the 80\,\% threshold by a red dashed line. (a) PL signal as a function of time when active stabilisation is off. (b) Spatially-resolved PL scan of the SPS. (c) PL signal as a function of the distance between the fibre and the sample surface.}
    \label{feature_norm}
\end{figure}

For practical use in quantum communication, a stable experimental setup is required over several hours. An ultra-stable manual 3-axis nanopositioning bench has been used for holding the fibre while the sample is fixed on an XYZ piezo stage. Thanks to this combination, we managed passively to maintain at least 80\% of the signal for 100 minutes (figure~\ref{feature_norm}a). This typical timescale thus allows, whenever necessary, to implement active stabilization with small feedback bandwidth over a few minutes.

The aforementioned spatial stabilization has three spatial components, two along the sample surface plane (X and Y), and a third one along the sample-to-fibre distance (i.e. along the Z optical axis of the system). The numerical aperture of the current fibre, 0.42, allows us to obtain a spot size of 2.2\,$\mu$m (figure~\ref{feature_norm}b), comparable to previously published results using microscope objective-based setups~[12]. When the position of the fibre is varied along the optical axis (i.e. the height of the fibre with respect to the sample surface), the optimum focal point for collection and its FWHM can be compared to the Rayleigh length, which has been calculated for Gaussian beams at 4.6\,$\mu$m. The experimental curve (figure~\ref{feature_norm}c) reproduces the numerical estimation and indicates that 80\% of the signal should be retained across a 3.0\,$\mu$m distance along the optical axis.

A comprehensive evaluation of the existing figures of merit reveals that an all-fibre configuration exhibits a level of performance that is comparable to that of conventional free-space and microscope-objective-based setups. Furthermore, it is evident that mechanical stability in a plane perpendicular to the optical axis is of greater significance than stability along the optical axis direction.

\section{Quantum characterization}

\subsection{Purity measurement} \label{purity meseaurement}
As mentioned earlier, a crucial characteristic of SPS for establishing QKD protocols is their single-photon purity, which affects the risks of splitting attacks, increasing thereby the security of the keys.

\begin{figure}
    \centering
    \includegraphics[width=1\linewidth]{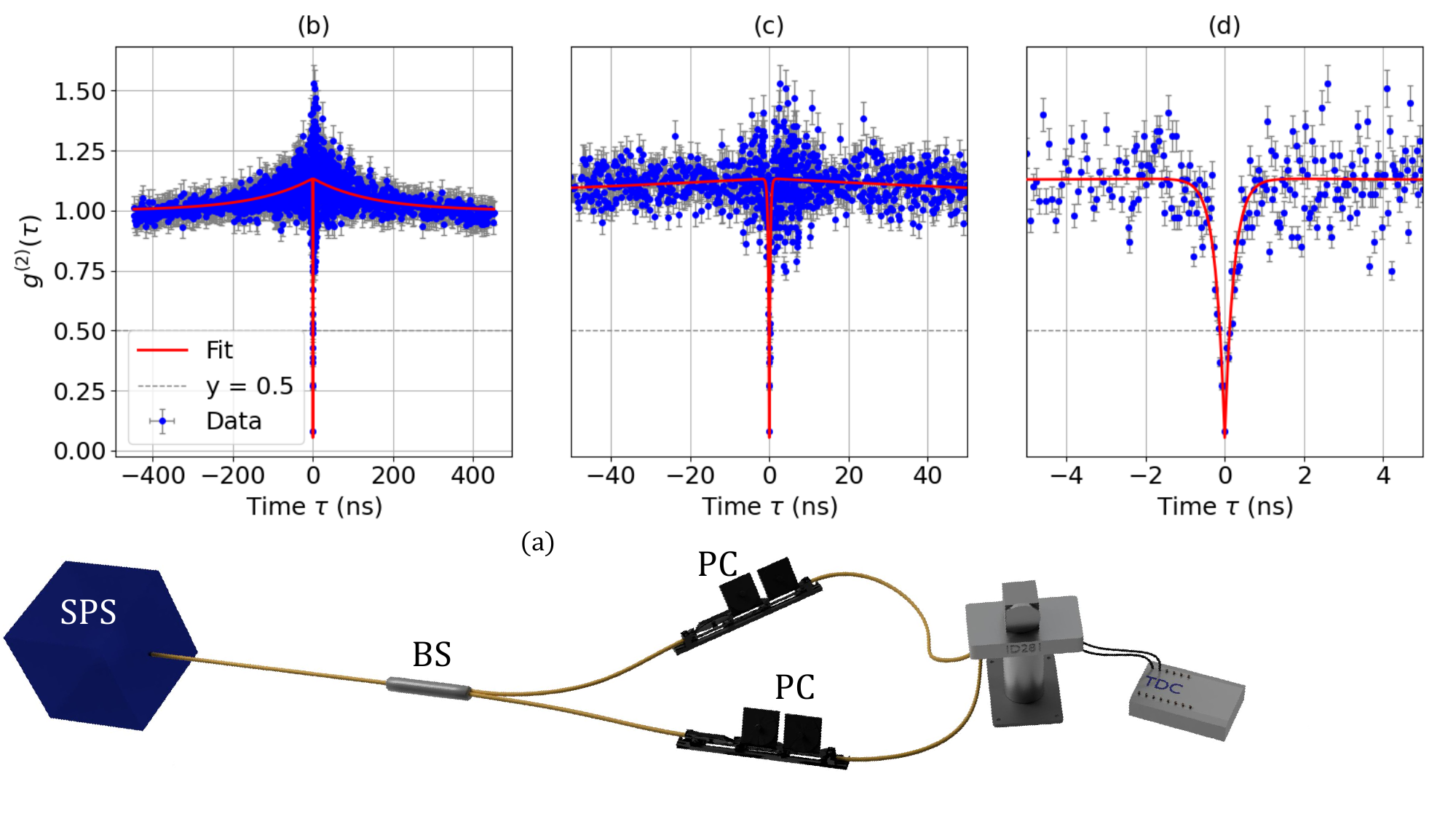}
    \caption{\centering Single photon emitter purity measurement. (a) Experimental setup of an HBT interferometer to measure the autocorrelation function. SPS : single photon source, BS : beam splitter and PC : polarized controller. (b), (c) \& (d) Measurement of intensity autocorrelation with a pump power of 1 mW. The data are shown in blue with their associated error bars, and the fitting curve, following the equation $g^{(2)}(\tau) = 1 - \beta_1 e^{-\tau / \tau_1} + \beta_2 e^{-\tau / \tau_2}$, is shown in red. The measurement lasted for 250\,minutes, with a sampling time resolution \( \delta t \) of 40\,ps and 25,000 photons detected per second per mW. For proper normalization purpose, we show the time delay up 500\,ns in (b) and show in (c) \& (d) a zoom-in around zero-delay time.}
    \label{g2}
\end{figure}

The purity of the emitted photons is evaluated by mean of a second-order autocorrelation measurement, referred to as \( g^{(2)}(\tau)\), using an all-fibred Hanbury-Brown and Twiss (HBT) setup, schematically shown in figure~\ref{g2}a. The value of \(g^{(2)}(\tau)\) at zero time delay is proportional to the probability of emitting two photons simultaneously. Experimentally, $g^{(2)}(0)$  is evaluated using the following equation:
\begin{equation}
g^{(2)}(\tau) = \frac{p_{12}(\tau)}{p_1 \cdot p_2}
\end{equation}
with \( p_{12}(\tau) \) the probability associated to a two-photon event for a time delay $\tau$ (coincidence detections on detectors 1 and 2), normalized by the product of the single photon probability \( p_1 \), \( p_2 \) per measurement time bin. In practice, it is defined by the resolution ($\delta t$) of the time-tagging device and the formula is calculated as follow :
\begin{equation}
g^{(2)}(\tau) = \frac{N_{12}(\tau)}{N_1 \cdot N_2 \cdot\delta t}
\end{equation}
with \( N_{12}(\tau) \) the coincidence detections rate per time bin $\delta t$ , and \( N_1 \), \( N_2 \) the single photon rate at a time delay $\tau$.

The measurement, shown in figure~\ref{g2}, has been performed at room temperature, for a pumping power of 1\,mW while active stabilization is enabled for more than 3 hours. Figure~\ref{g2}b presents the measurement over a large time interval of 1000\,ns necessary for accurate normalisation of the \( g^{(2)}(\tau)\), while in figure~\ref{g2}c and 4d a zoom-in around the antibunching dip is shown. The \( g^{(2)}(\tau)\) curve has been fitted using the following function~\cite{patel2022probing}:

\begin{equation}
    g^{(2)}(\tau) = 1 - \beta_1 e^{-\tau / \tau_1} + \beta_2 e^{-\tau / \tau_2}
\end{equation}

where \( \tau_1 \) represents the antibunching characteristic time, associated with (but not given directly by) the lifetime of the level responsible for the emission of single photons, and \( \tau_2 \) is a characteristic bunching time associated with transitions between our emissive state and additional metastable states, while \( \beta_1 \) and \( \beta_2 \) represent the respective amplitudes. This function, with a single antibunching and bunching characteristic times, is typically used to fit the autocorrelation functions of three-level systems~\cite{fishman2023photon} and is well-suited for the current emitter~\cite{meunier2023telecom}.

We obtain \( g^{(2)}(0) = (5.9 \pm 0.5) \times 10^{-2} \) and an antibunching lifetime \( \tau_1 \) of 230\,ps. This value of \( g^{(2)}(0) \) reproduces the state-of-the-art for room temperature and telecom wavelength SPS, and the lifetime is consistent with measurements already performed in reference~\cite{mu2021room}.

\subsection{Purity $\&$ brightness}
\begin{figure}
    \centering
        \centering
        \includegraphics[width=\linewidth]{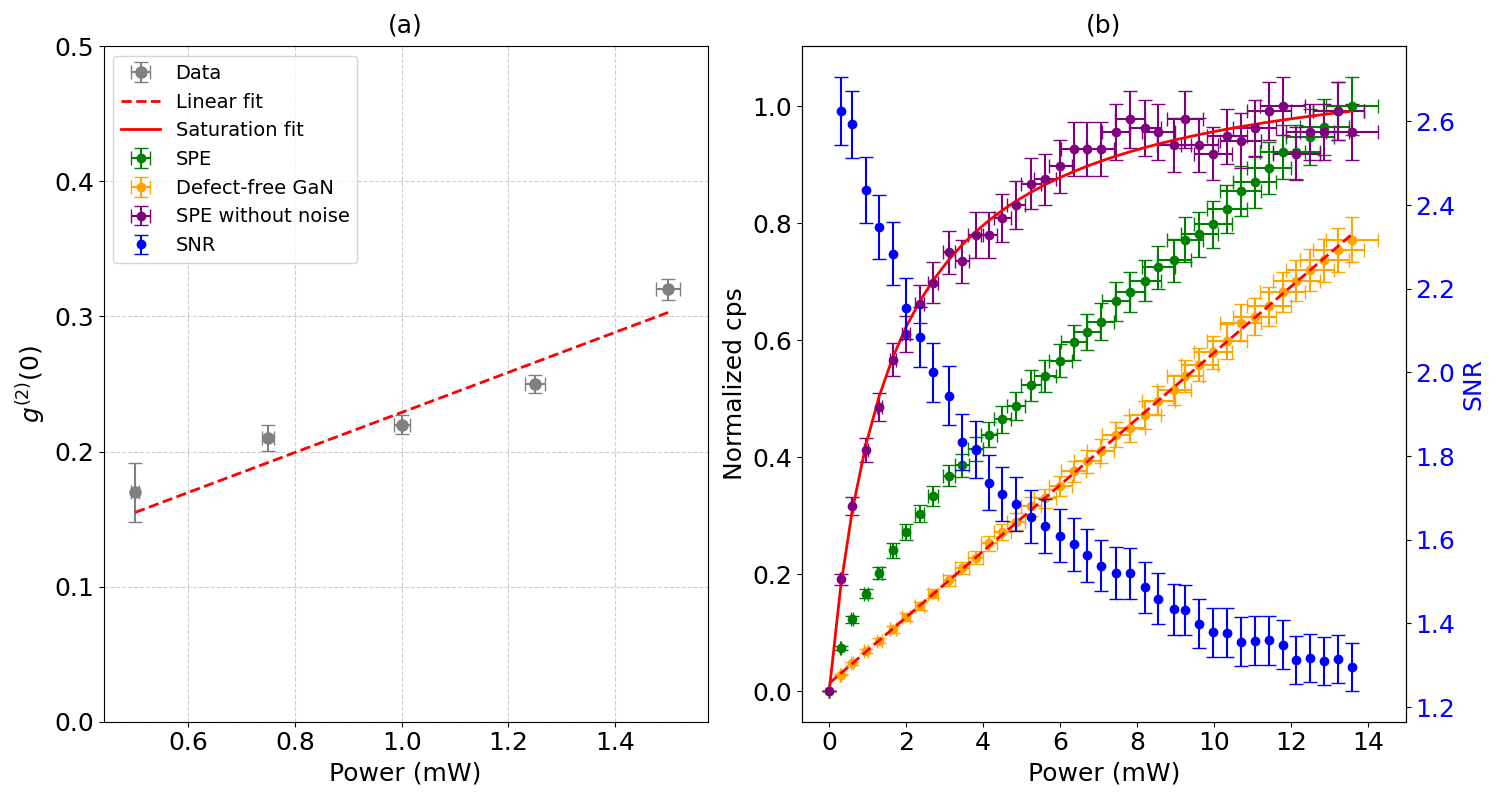}
        \label{g2mw}    
    \caption{\centering Comparaison of purity and brightness versus pump power. (a) $g^{(2)}(0)$ as a function of pump power. The measurement was conducted with an integration time of 90 minutes and with the same experimental setup. (b) Normalised counts per second and signal-to-noise ratio as a function of pump power. The green, yellow and purple dots represent normalised emission from the SPE, a SPS-free GaN location and single photon number in the absence of noise (removing the contribution from the GN background), respectively. The blue curve represents the signal-to-noise ratio. The red lines represent the different regressions: linear for SPS-free GaN location and saturation for single photons without noise following the equation $I = I_{sat} \frac{P}{P_{sat} + P}$.}
    \label{power}
\end{figure}

While the purity of the SPS is a major factor to be taken into account for secured key exchange and distillation, the brightness of the emitter is also of paramount importance from the practical point of view. Thus, a tradeoff might be found between purity and emission rate, given that both (figure~\ref{power}) depend on the pumping power.

To assess quantitatively this tradeoff between purity and emission rate, we plotted $g^{(2)}(0)$ as a function of pump power (Figure~\ref{power}a). It should be noted that study has not been carried out under the same conditions as that of \ref{purity meseaurement} section. For figure~\ref{power}a, we have used fixed conditions for each pump power, with a shortened measurement time of 90 minutes and no active stabilisation. This explains the difference of purity reported figure~\ref{g2}a with a measurement time of 250\,min, a dark count level reduced to a minimum and active stabilisation. We can see that $g^{(2)}(0)$ increases linearly with pump power. This phenomenon is explained by the optical noise from the fluorescence of GaN, growing linearly at a faster rate than single photon the spontaneous emission.Showing typical saturation behavior related to the lifetime. This is confirmed in figure~\ref{power}b, where one can observe the behavior of the SPS intensity (green dots) as a function of pump power is simply the sum of the behavior of GaN fluorescence (yellow dots), which is linear in this power range, and that of the single photons (purple dots), which saturates after a few milliwatts. Note that to highlight the behavior of the SPS emission saturation, all three curves are normalized, but in absolute intensity the green curve is the sum of the yellow and purple dots. The saturation of the SPS emission can be adjusted (red line) as $I = I_{sat} \frac{P}{P_{sat} + P}$. Furthermore, it's interesting to plot indicating that the signal-to-noise ratio (blue dots) as a function of pump power, the fluorescence of GaN grows without saturation, contrary to the SPS. Consequently, it becomes essential to identify the operation conditions respecting the requisite level of purity to implement a QKD protocol between two parties while maximising the total brightness.

\subsection{Quantum state tomography}
Polarisation and phase-time encoding are usually used for QKD protocols. Phase-time encoding relies on the coherence properties of the photon wave packet and the temporal resolution of the detectors to ensure the security of the protocols. This approach is particularly advantageous over fibre links where dynamical birefringence, induced by environmental disturbance, are too detrimental. However, this method requires high-speed modulators and remote interferometric setups to be stabilized, which can increase the technical complexity of its implementation~\cite{zhang2024metropolitan}.

Polarization encoding, on the other hand, is generally appreciated for its ease of experimental implementation, facilitated by the availability of standard optical devices such as wave-plates and polarizers. The downside of polarisation encoding is the requirement for active birefringence compensation and a highly polarized source of photonic states~\cite{zhang2024polarization}.

\begin{figure}
    \centering
    \includegraphics[width=0.9\linewidth]{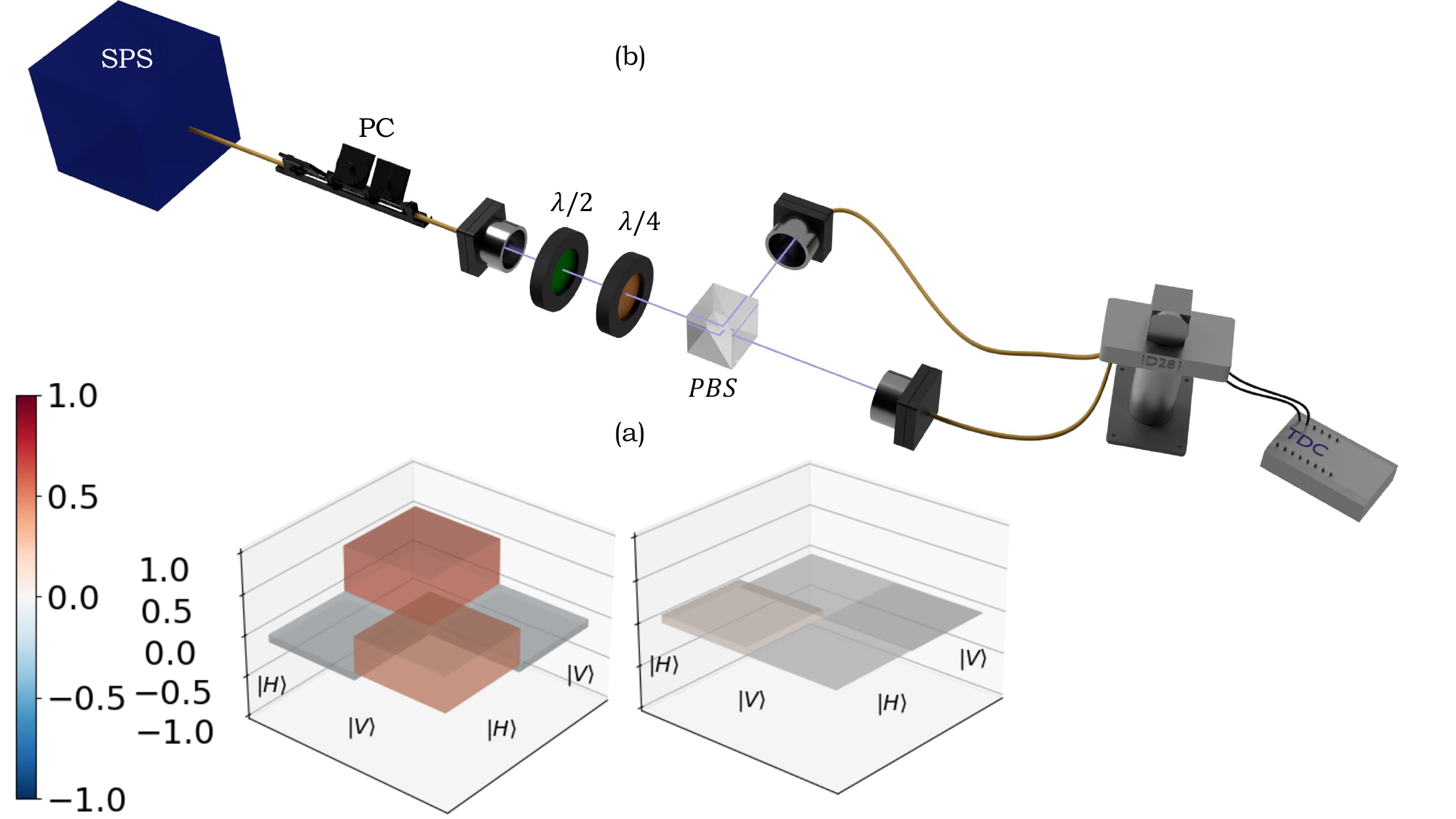}
    \caption{\centering Measurement of the polarisation state of the single photons. (a) Experimental setup of a tomographic bench. $\lambda$/2 and $\lambda$/4 : half-wave and quarter-wave plate. SNSPD : superconducting nanowire single-photon detector and PBS : Polarized beam splitter. (b) Tomographic measurement of the SPS. The matrix on the left represents the real part, while the matrix on the right represents the imaginary part. Calculation of the eigenvalue shows a partially polarisation emission of 65\%.}
    \label{tomo}
\end{figure}

Tomographic measurements have been used to assess the polarization state of the single photons emitted by our SPS and validate its appropriateness for QKD protocols exploiting polarization as an observable. We reconstruct the corresponding density matrix in figure~\ref{tomo}a, which characterizes the quantum state of the emitted photons.

The current emitter is weakly polarized (figure~\ref{tomo}b), since the density matrix eigenvalues are respectively equal to 0.35 and 0.65. This means that if QKD was performed on the polarization observable, 35\% of the photons would have to be sacrificed in order to be able to share keys. Thus, for this emitter phase-time, would certainly be more appropriate~\cite{zhang2024metropolitan}.

 \section{Conclusion}
We have investigated, thanks to an all-fibred micro photoluminescense system, the properties of GaN point defects emitting single photons in the telecom O-band. We have shown the potential of an all-fibred setup to realize a compact and reliable room temperature single photon source for real-field quantum communication. We have exploited a lensed fibre enabling high collection efficiency, with a state-of-the-art purity $g^{(2)}(0) = (5.9 \pm 0.5) \times 10^{-2}$ and a practical emission rate of up to 60\,kHz. We also have shown that purity and SNR is mainly limited by fluorescence of GaN and that fibred setup doesn't add any significant noise. The fully fibred setup is mechanically stable for several hours. An improved setup, implementing photonic structures around the emitters for enhancing light extraction, should allow us to integrate this source into a compact package for practical deployment, meeting the basic requirements for establishing a QKD protocol with a purity and emission rate beyond the state-of-the-art.

\begin{backmatter}
\bmsection{Funding}
This work has been conducted within the framework of the French government financial support managed by the Agence Nationale de la Recherche (ANR), within its Investments for the Future programme, under the Université Côte d’Azur UCA-JEDI project (Quantum@UCA, ANR-15-IDEX-01), under the Stratégie Nationale Quantique through the PEPR QCOMMTESTBED project (ANR 22-PETQ-0011) and also under the France 2030 project CIPHER-Q (ANR-22-CE47-0006-01). This work has also been conducted within the framework of the OPTIMAL project, funded by the European Union and the Conseil Régional SUD-PACA by means of the ‘Fonds Européens de développement regional’ (FEDER). The authors also acknowledge financial support from the European comission, project 101114043-QSNP, and project 101091675-FranceQCI.

\bmsection{Acknowledgment}
J.Z.P has also received financial support from the CNRS through the 80| Prime program and from CNRS and NTU through the CNRS-NTU Excellence Science Joint Research Program (UNIQUE).

N.L would like to express his gratitude to EURSPECTRUM and PEPR Qcommtestbed 22-PETQ-0011 for their financial support for his thesis.
\bmsection{Disclosures}
The authors declare no conflicts of interest.

\bmsection{Data Availability Statement}
The data that support the findings of this study are available upon reasonable request to the authors.

\end{backmatter}

%%%%%%%%%%%%%%%%%%%%%%% References %%%%%%%%%%%%%%%%%%%%%%%%%

%%%%%%%%%% If using BibTeX:
\bibliography{biblio}

%%%%%%%%%% If preparing manually:
% \begin{thebibliography}{1}
% \newcommand{\enquote}[1]{``#1''}

% \bibitem{Zhang:14}
% Y.~Zhang, S.~Qiao, L.~Sun, Q.~W. Shi, W.~Huang, L.~Li, and Z.~Yang,
%   \enquote{Photoinduced active terahertz metamaterials with nanostructured
%   vanadium dioxide film deposited by sol-gel method,}
%   {\protect\JournalTitle{Optics Express}} \textbf{22}, 11070--11078 (2014).

% \bibitem{Optica}
% {Optica}, \enquote{{Optica Publishing Group},}
%   \url{http://www.opg.optica.org}.

% \bibitem{FORSTER2007}
% P.~Forster, V.~Ramaswamy, P.~Artaxo, T.~Bernsten, R.~Betts, D.~Fahey,
%   J.~Haywood, J.~Lean, D.~Lowe, G.~Myhre, J.~Nganga, R.~Prinn, G.~Raga,
%   M.~Schulz, and R.~V. Dorland, \enquote{Changes in atmospheric consituents and
%   in radiative forcing,} in \enquote{Climate Change 2007: The Physical Science
%   Basis. Contribution of Working Group 1 to the Fourth Assesment Report of
%   Intergovernmental Panel on Climate Change,}  S.~Solomon, D.~Qin, M.~Manning,
%   Z.~Chen, M.~Marquis, K.~B. Averyt, M.~Tignor, and H.~L. Miler, eds.
%   (Cambridge University Press, 2007).

% \end{thebibliography}

\end{document}